\documentclass[a4paper]{spie}  %>>> use this instead for A4 paper
\usepackage[]{graphicx}

\title{The VORTEX project: first results and perspectives} 

\author{Olivier Absil\supit{a}, Dimitri Mawet\supit{b}, Christian Delacroix\supit{a}, Pontus Forsberg\supit{c}, Mikael Karlsson\supit{c}, Serge Habraken\supit{a}, Jean Surdej\supit{a}, Pierre-Antoine Absil\supit{d}, Brunella Carlomagno\supit{a}, \\ Valentin Christiaens\supit{e,a}, Denis Defr\`ere\supit{f}, Carlos Gomez Gonzalez\supit{a}, Elsa Huby\supit{a}, A\"issa Jolivet\supit{a}, Julien Milli\supit{b}, Pierre Piron\supit{a}, Ernesto Vargas Catalan\supit{c}, Marc Van Droogenbroeck\supit{g}
\skiplinehalf
\supit{a}Department of Astrophysics, Geophysics and Oceanography, University of Li\`ege, \\ 17 all\'ee du Six Ao\^ut, B-4000 Sart Tilman, Belgium; \\
\supit{b}European Southern Observatory, Alonso de C\'ordova 3107, Vitacura, Santiago, Chile; \\
\supit{c}\AA{}ngstr\"om Laboratory, Uppsala University, L\"agerhyddsv\"agen 1, SE-751 21 Uppsala, Sweden; \\
\supit{d}Department of Mathematical Engineering, UCLouvain, B-1348 Louvain-la-Neuve, Belgium; \\
\supit{e}Departamento de Astronom\'ia, Universidad de Chile, Casilla 36-D, Santiago, Chile; \\
\supit{f}Steward Observatory, University of Arizona, 633 N. Cherry Avenue, 85721 Tucson, USA; \\
\supit{g}Montefiore Institute, University of Li\`ege, B-4000 Sart Tilman, Belgium
}

\authorinfo{Send correspondence to O.A. (email: absil@astro.ulg.ac.be).}
\begin{document} 
  \maketitle 

%%%%%%%%%%%%%%%%%%%%%%%%%%%%%%%%%%%%%%%%%%%%%%%%%%%%%%%%%%%%% 
\begin{abstract}
Vortex coronagraphs are among the most promising solutions to perform high contrast imaging at small angular separations from bright stars. They feature a very small inner working angle (down to the diffraction limit of the telescope), a clear 360 degree discovery space, have demonstrated very high contrast capabilities, are easy to implement on high-contrast imaging instruments, and have already been extensively tested on the sky. Since 2005, we have been designing, developing and testing an implementation of the charge-2 vector vortex phase mask based on concentric subwavelength gratings, referred to as the Annular Groove Phase Mask (AGPM). Science-grade mid-infrared AGPMs were produced in 2012 for the first time, using plasma etching on synthetic diamond substrates. They have been validated on a coronagraphic test bench, showing broadband peak rejection up to 500:1 in the L band, which translates into a raw contrast of about $6\times 10^{-5}$ at $2 \lambda/D$. Three of them have now been installed on world-leading diffraction-limited infrared cameras, namely VLT/NACO, VLT/VISIR and LBT/LMIRCam. During the science verification observations with our L-band AGPM on NACO, we observed the beta Pictoris system and obtained unprecedented sensitivity limits to planetary companions down to the diffraction limit ($0.1''$). More recently, we obtained new images of the HR 8799 system at L band during the AGPM first light on LMIRCam. After reviewing these first results obtained with mid-infrared AGPMs, we will discuss the short- and mid-term goals of the on-going VORTEX project, which aims to improve the performance of our vortex phase masks for future applications on second-generation high-contrast imagers and on future extremely large telescopes (ELTs). In particular, we will briefly describe our current efforts to improve the manufacturing of mid-infrared AGPMs, to push their operation to shorter wavelengths, and to provide deeper starlight extinction by creating new designs for higher topological charge vortices. Within the VORTEX project, we also plan to develop new image processing techniques tailored to coronagraphic images, and to study some pre- and post-coronagraphic concepts adapted to the vortex coronagraph in order to reduce scattered starlight in the final images. 
\end{abstract}

%>>>> Include a list of keywords after the abstract 


%%%%%%%%%%%%%%%%%%%%%%%%%%%%%%%%%%%%%%%%%%%%%%%%%%%%%%%%%%%%%
\section{Introduction}
\label{sec:intro}  % \label{} allows reference to this section

The last few years have seen tremendous progress in the field of exoplanet imaging, with the first resolved images of several giant planets around nearby main sequence stars \cite{Marois08,Kalas08,Lagrange10,Kuzuhara13,Rameau13,Bailey14}. These discoveries have been enabled mostly by the optimization of observing strategies and by recent progress in image processing algorithms over the last decade. Despite all these efforts, the current images still generally concern relatively wide companions ($\ge 0.5''$ from the star) at moderate ($\sim 10$ mag) to high ($\sim 15$ mag) contrasts. Achieving high contrast at short angular separations requires improvement in the hardware, in particular the combination of an excellent wavefront control with an appropriate starlight cancellation method working over a wide spectral band. The vortex coronagraph is among the most promising solutions in that context, as it enables imaging down to the diffraction limit of the telescope and can be made achromatic over large bandwidths (see e.g.\ [\citenum{Mawet11,Mawet12}] for recent reviews).

In 2005, our team proposed for the first time to use an optical vortex as a focal-plane phase mask for coronagraphic applications \cite{Mawet05}. Since then, we have been developing this original concept, based on the use of subwavelength gratings (SG) to synthesize the desired phase ramp around the optical axis. In this paper, we report on the first results obtained in this context and describe the main perspectives for the years to come regarding the scientific exploitation, the continuous improvement of the technology and the optimization of the systems aspects.

%%%%%%%%%%%%%%%%%%%%%%%%%%%%%%%%%%%%%%%%%%%%%%%%%%%%%%%%%%%%%
\section{From concept to science-grade vector vortices}

Historically, our proposal to use a vortex phase mask derives from the need to provide an achromatic phase mask. We first proposed to use subwavelength gratings to produce achromatic half wave plates, which can be used to replace the stepped phase pattern of the Four Quadrant Phase Mask (FQPM) \cite{Mawet03,Mawet05b}. The phase shift then becomes ``vectorial'', as the two polarisations are treated in a separate way (with opposite phase shifts for the two incoming polarisations). 

We then figured out that we could get rid of the quadrant transitions (which are blind to off-axis companions) by making the phase shift continuous, instead of discrete. A straightforward solution was found with subwavelength gratings, which just need to be made circular around the optical axis to produce the appropriate phase ramp, where the phase grows from 0 to $4\pi$ on a complete revolution around the center. Due to its particular geometry, we nicknamed this concept the Annular Groove Phase Mask (AGPM) \cite{Mawet05}. In a more general way, we refer to it as the Subwavelength Grating Vortex Coronagraph of topological charge 2 (SGVC2), where the topological charge is defined as the (even) number of times the geometric phase accumulates $2\pi$ along a closed path surrounding the central phase singularity \cite{Mawet05,Delacroix14}.

After some preliminary attempts on silicon, we decided to use diamond as a substrate to etch our subwavelength gratings. Diamond has many assets that fully compensate for the manufacturing challenge: transparency from the ultraviolet to the microwave regime, high refractive index (around 2.4 in the near-infrared, which reduces the requirement on grating aspect ratio), toughness, hardness, thermal conductivity, electrical insulation, mechanical and chemical stability, etc. Furthermore, the \AA{}ngstr\"om Laboratory at Uppsala University has specialized in the etching of micro-structures on synthetic diamond substrates, which naturally led to a privileged partnership between Uppsala and Li\`ege.

\begin{figure*}[!t]
\begin{center}
 \includegraphics[scale=0.35]{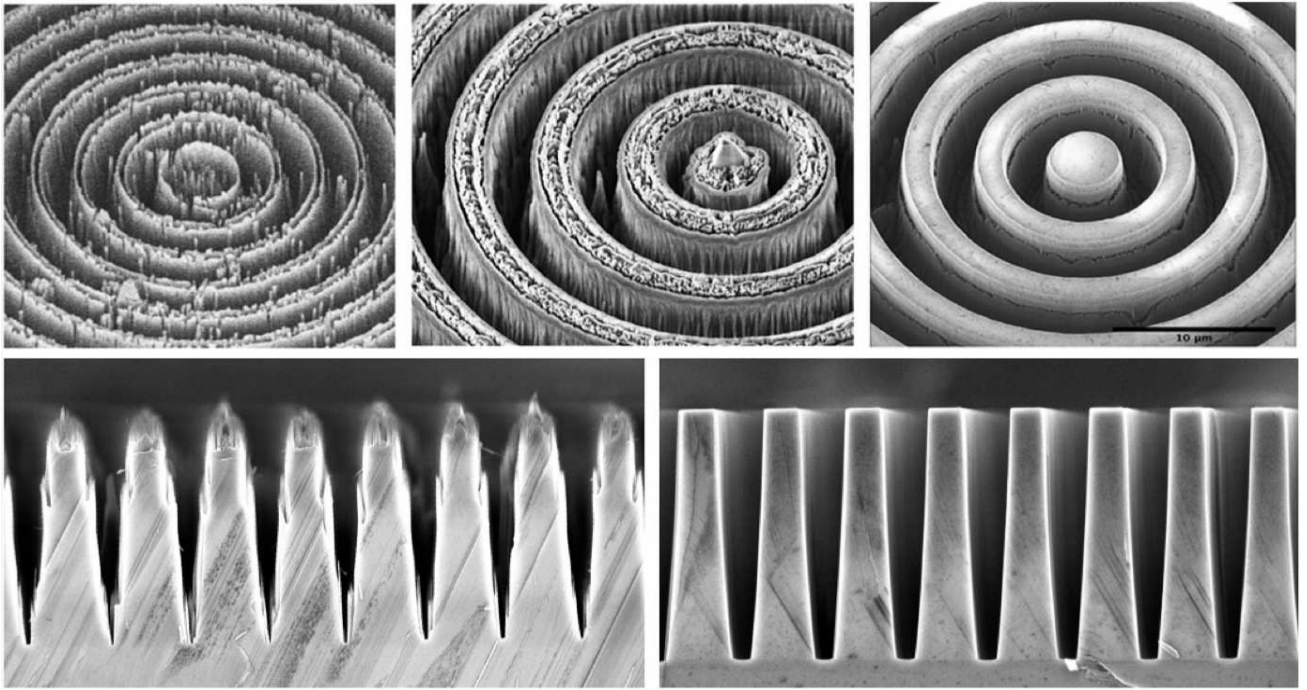}
\end{center}
\caption{Evolution of the manufacturing of diamond AGPMs. Top: N-band AGPMs, with periods $\sim4.6~\mu$m, manufactured in November 2009, October 2010, and February 2012 (from left to right). Bottom: cracked spares of L-band AGPMs, with periods $\sim1.4~\mu$m, manufactured in March 2011 and September 2012 (from left to right). Taken from [\citenum{Delacroix13b}].}
 \label{fig:agpm}
\end{figure*}

While first etching tests were carried out in 2005 (in the context of the FQPM made of subwavelength gratings), the first attempt at etching full-blown AGPMs dates back to 2009. Mastering the process of etching the relatively deep grooves (a few micrometer deep) of the AGPM onto a diamond substrate took three more years (see [\citenum{Karlsson10,Forsberg13}] for details), resulting in the first science-grade AGPMs in 2012 (see Fig.~\ref{fig:agpm}). Because it relaxes the requirement on the etching period, the mid-infrared region was chosen for the first AGPMs, targeting the L and N transmission bands of the atmosphere (respectively ranging from 3.5 to 4.1 $\mu$m and from 8 to 13 $\mu$m, which leads to periods of about 1.4 and 4 $\mu$m). L-band AGPMs were tested on an infrared coronagraphic test bench at Observatoire de Paris, resulting in broadband peak rejection ratios up to 500:1\cite{Delacroix13}, which translates into a raw contrast of about $6\times 10^{-5}$ at $2 \lambda/D$. Due to the lack of cryogenic coronagraphic test bench available to us, N-band AGPMs have not yet been fully characterized.

Based on these first successes, the project can now move on to a new phase, as funding was secured in 2013 from an ERC Starting Grant (PI: Absil) and an ARC grant from the University of Li\`ege (PI: Surdej). In the next sections, we describe the main goals and first results of these two coordinated projects (which, taken together, are referred to as the VORTEX project).

%%%%%%%%%%%%%%%%%%%%%%%%%%%%%%%%%%%%%%%%%%%%%%%%%%%%%%%%%%%%%
\section{First light and on-going scientific exploitation}

After validating our mid-infrared AGPMs in the lab, the next step was to install them at the telescope. This was performed in 2012-2013 on three different infrared cameras installed at world-leading observatories: VISIR and NACO at the Very Large Telescope, and LMIRCam at the Large Binocular Telescope.

The installation of an N-band AGPM on VISIR was part of the major upgrade that VISIR underwent in 2012, under the leadership of ESO and CEA Saclay. Although the performance of our N-band AGPMs had never been validated on a coronagraphic test bench, we were confident that it would provide a peak rejection of at least a few tens, based on Rigorous Coupled Wave Analysis simulations using the measured grating parameters. These measurements were performed with Scanning Electron Microscopy (SEM, Fig.~\ref{fig:agpm}), both on the surface of the science-grade component and after cleaving several test components to precisely measure their grating parameter (including the depth and sidewall angle, which is not easy to evaluate from surface SEM due to the depth of the grooves). The AGPM hit the sky in mid-2012 during the commissioning of VISIR, but has not been fully tested so far due to the interruption of the VISIR commissioning. Preliminary results obtained on internal sources suggest that the N-band AGPM lives up to our expectations in terms of peak rejection. We are now waiting for the re-commissioning of VISIR to perform a full validation of the N-band AGPM performance.

\begin{figure*}[!t]
\begin{center}
 \includegraphics[scale=0.43]{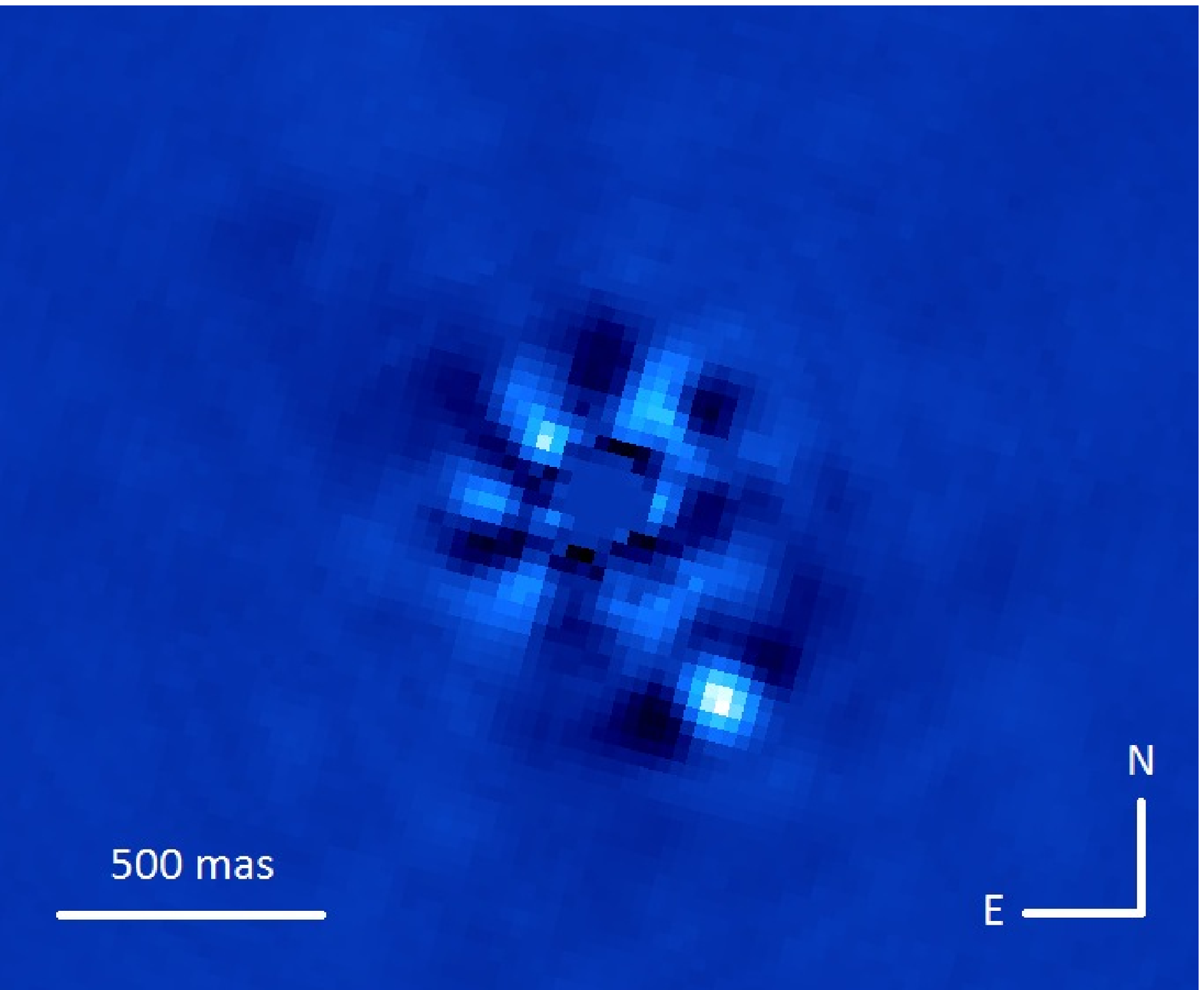} \quad
 \includegraphics[scale=0.5]{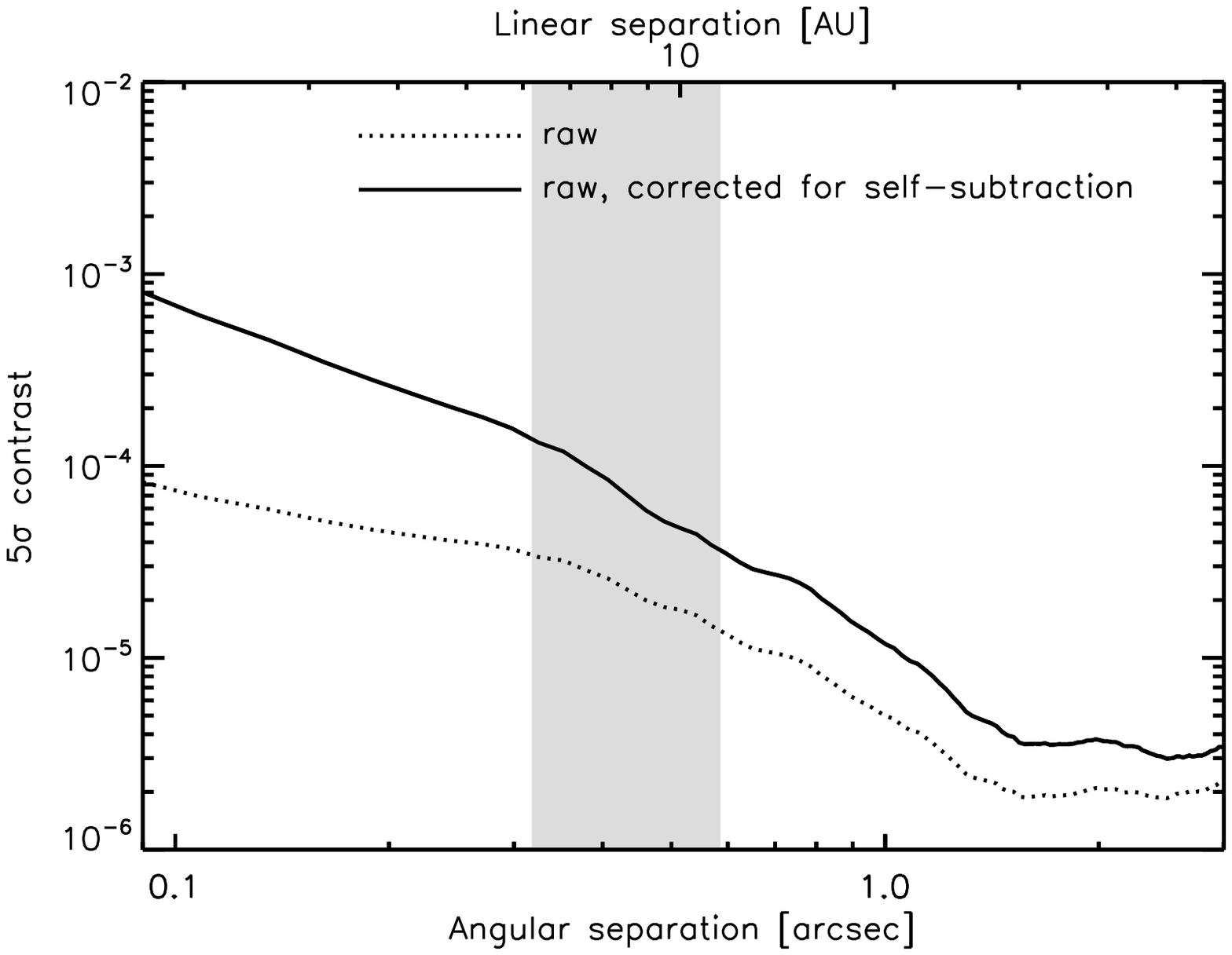}
\end{center}
\caption{Left: final image of $\beta$ Pic and its giant planet, after PCA-ADI processing and correction for the throughput of the PCA-ADI processing. Right: sensitivity to companions expressed in terms of contrast ($5\sigma$) as a function of angular separation. The gray shaded region represents the region where the sensitivity estimation is affected by the presence of $\beta$ Pic b. Due to small number statistics, the confidence level (not represented here) associated to our sensitivity limits degrades significantly at small angular separations \cite{Mawet14}. Adapted from [\citenum{Absil13}].}
 \label{fig:betpic}
\end{figure*}

An L-band AGPM was installed on NACO after being validated on the coronagraphic test bench of Observatoire de Paris. It obtained first light in December 2012, resulting in the serendipitous discovery of a late-type stellar companion at only two beamwidths from the F0-type main sequence star HD~4691 (magnitudes $V=6.8$, $L=5.9$) \cite{Mawet13}. These first observations showed an instantaneous on-sky peak rejection of about 50 for the AGPM on NACO, and demonstrated the great potential of the AGPM to address companions at very short angular separation from bright objects. Science verification observations were then obtained on the night of 31 January 2013, during which beta Pictoris was observed for about 3.5 hours. These observations allowed us to search for planetary-mass companions down to only 2~AU from the star (see Fig.~\ref{fig:betpic}), thereby significantly improving upon previous works \cite{Absil13}. This data set also provided new constraints on the debris disk morphology at short angular separation \cite{Milli14}.

Finally, an L-band AGPM was installed in 2013 on LMIRCam inside the LBTI, the imaging and interferometric instrument for the LBT developed by the University of Arizona. First light was successfully obtained in October 2013. These observations are reported in another paper within these proceedings \cite{Defrere14}.

Because they provide access to unprecedentedly small angular separations (in terms of beamwidth), the first scientific observations carried out with our mid-infrared AGPMs have raised an important question regarding the very definition of sensitivity to off-axis companions. Indeed, the sensitivity is generally defined with respect to the ``local'' noise in the image, computed as the root mean square of the speckle intensity in concentric annuli around the star. At an angular distance of $1\lambda/D$ (resp.\ $2\lambda/D$) from the center, the number of independent resolution elements (i.e., speckles) in an annulus only amounts to six (resp.\ twelve). We are thus heavily limited by small number statistics, which changes the confidence level associated to a given signal-to-noise ratio. This issue is investigated in depth in a forthcoming paper \cite{Mawet14}, where specific recommendations are made on the computation of sensitivity limits at very small separations.

We are now in a status where the scientific exploitation of the AGPM has become a reality, with on-going programs on NACO and LMIRCam. High-contrast L-band observations will be particularly useful in the coming years to follow up the detection of new planets with second-generation near-infrared high-contrast imagers such as Gemini/GPI, VLT/SPHERE, or Subaru/SCExAO. Collecting broadband photometry at mid-infrared wavelengths for the planets detected in the near-infrared will be particularly useful to constrain models (and most notably the cloud structure). We also plan to carry out several dedicated scientific programs, such as a small survey of Jupiter-sized planets around young, nearby late-type dwarfs. To get the most of these observations, we are currently developing a custom image processing pipeline, based on a principal component analysis (PCA) of the individual frames to further reduce the residual starlight from the final image \cite{Absil13}. More advanced versions of PCA-based processing and new algorithms are currently being tested to improve upon the state of the art. The pipeline is currently optimized for angular differential imaging data sets, but will be extended to reference-star differential imaging and spectral deconvolution in the coming years.

% say a few words about Keck?

%%%%%%%%%%%%%%%%%%%%%%%%%%%%%%%%%%%%%%%%%%%%%%%%%%%%%%%%%%%%%
\section{Beyond mid-infrared AGPMs: design and manufacturing}

Now that our first-generation mid-infrared AGPMs have been validated in the lab and on the sky, we are working on the improvement of vector vortex phase masks in three main ways: (i) continuous improvement of the manufacturing process to improve the performance of mid-infrared AGPMs, (ii) downscaling of the design and manufacturing to shorter wavelengths, and (iii) creation of new designs for vortex phase masks with higher topological charges. These various goals are the subject of other, more detailed papers within these proceedings \cite{Forsberg14,Carlomagno14,Delacroix14} and will only be shortly reviewed here.

First, the etching process is being improved to make it more stable and reproducible, as detailed in [\citenum{Forsberg14}]. This will allow us to reach the optimal grating parameters for mid-infrared AGPMs in a more consistent way, without needing to repeat the process several times to (serendipitously) end up with an optimal grating. In parallel, the optimal grating parameters are being re-evaluated based on the latest results of the manufacturing. This effort is geared up towards delivering optimized mid-infrared AGPMs for METIS, the future mid-infrared imager of the European Extremely Large Telescope (E-ELT) \cite{Brandl14}. The improvement of the etching recipe also concerns the anti-reflective (AR) gratings that we etch on the back side of the diamond substrates of our AGPMs. The AR grating is mandatory to reach the best possible peak starlight rejection, as multiple reflections within the substrates would lead to a spurious ghost signal. New designs beyond the simple one-step pattern used in our first generation AGPMs are being investigated \cite{Forsberg13}.

Based on this improved etching recipe, we expect to be able to etch even finer gratings, and thereby address shorter operating wavelengths. In particular, first etching tests are being performed for K-band gratings, which feature a grating period of about 800~nm (instead of about 1.4~$\mu$m in the L band). This leads to new challenges in the etching process, such as the redeposition of etched material on top of the grooves, which could stop the etching before a sufficient depth is reached. The non-vertical sidewalls may also become problematic for smaller grating periods. Here again, optimized grating parameters will be computed based on the first etching results. By producing AGPMs at shorter wavelengths, we could consider such phase masks as possible upgrades to state-of-the-art near-infrared imagers recently installed at world-leading observatories (e.g., Gemini/GPI, VLT/SPHERE, etc).

Finally, in the context of future near-infrared applications on the E-ELT, the finite size of bright stars may become a significant limitation to the performance of charge-2 vortex phase masks. We are therefore currently considering the extension of our subwavelength grating design to charge-4 (or even charge-6) vortices. Designing and etching subwavelength gratings for (even) topological charges higher than two is a major challenge, as the grating period is supposed to change drastically over the substrate as a function of the radial and azimuthal coordinates. New solutions to etch such gratings while preserving the subwavelength regime (so that only the zeroth order of the grating is propagated) are currently being investigated \cite{Delacroix14}. In addition to being mostly insensitive to the star size, such vortices would also be more resilient to pointing errors and low-order aberrations, which may reveal useful in the context of high-contrast instruments on extremely large telescopes \cite{Mawet10}. This however comes at a price, that is a degradation of the inner working angle, which increases from $0.9 \lambda/D$ for a charge-2 to $1.6 \lambda/D$ for a charge-4 vortex coronagraph.

%%%%%%%%%%%%%%%%%%%%%%%%%%%%%%%%%%%%%%%%%%%%%%%%%%%%%%%%%%%%%
\section{Performance testing and coronagraphic system aspects}

To enable the in-house testing of our near- and mid-infrared vortex phase masks, we are currently building at the University of Li\`ege an infrared coronagraphic test bench, nicknamed VODCA (VORTEX Optical Demonstrator for Coronagraphic Applications). It will operate at wavelengths ranging from 1 to 5~$\mu$m. The test bench is currently under construction, and is described in another publication within these proceedings \cite{Jolivet14}. It will feature a broadband infrared super continuum laser source, an all-reflective design, and a commercial infrared cooled InSb camera.

Beyond the testing of infrared vortex phase masks, VODCA will be used to develop and validate various concepts and techniques to enhance the coronagraphic performance at the system level, such as:
\begin{itemize}
\item \textit{Apodization.} Introducing a ring-shaped amplitude apodization in an upstream pupil plane can mitigate the diffraction from a central obscuration in a standard telescope \cite{Mawet13b,Mawet14b}. We plan to test this concept on our bench by introducing both a central obscuration and an apodizer in the input pupil of the coronagraphic bench.
\item \textit{Low-order wavefront sensing.} Various approaches have been proposed to measure the residual low-order optical aberrations affecting the coronagraph, targeting in particular the slowly evolving non-common path aberrations between the wavefront sensor and the coronagraph. We plan to test some of these approaches, such as the evaluation of tip-tilt from the images recorded by the scientific camera \cite{Mas12} or the use of a reflective Lyot stop to divert the stellar light to a specific wavefront analyzer \cite{Singh14}. We plan to adapt our bench in the future to install a low-order deformable mirror, which would be used to correct for these aberrations and provide an improved wavefront quality for testing our phase masks.
\item \textit{Post-coronagraphic solutions.} We are currently investigating new ways to improve the coronagraphic attenuation using post-coronagraphic devices. One of these options is based on the quantum properties of the light passing through a vortex phase mask.
\end{itemize}

%%%%%%%%%%%%%%%%%%%%%%%%%%%%%%%%%%%%%%%%%%%%%%%%%%%%%%%%%%%%%
\section{Conclusions}

Since its original proposal in 2005, the subwavelength grating vortex coronagraph of topological charge 2 (SGVC2, aka the Annular Groove Phase Mask) has been manufactured, tested in the lab and validated on the sky with various diffraction-limited infrared cameras. It has also provided its first scientific results, giving access to very small angular separations and thus urging us to reconsider the very definition of companion detection limits at small separation, which will become more and more critical as next-generation high-contrast imaging instruments come online. We are now starting a new phase of the VORTEX project, where we build upon this success to develop new SGVC with improved performance. These new vortex phase masks will be specifically designed for future applications on extremely large telescopes, such as the E-ELT. In parallel, we are currently developing new image processing algorithms for the optimal scientific exploitation of these coronagraphs, and we plan to test in the lab a few techniques to improve the coronagraphic performance with new hardware and software solutions. Taken together, these developments should allow fainter planets to be detected and characterized at smaller angular separations from bright nearby stars.

%%%%%%%%%%%%%%%%%%%%%%%%%%%%%%%%%%%%%%%%%%%%%%%%%%%%%%%%%%%%%
\acknowledgments     %>>>> equivalent to \section*{ACKNOWLEDGMENTS}       
 
The research leading to these results has received funding from the European Research Council under the European Union's Seventh Framework Programme (ERC Grant Agreement n.337569) and from the French Community of Belgium through an ARC grant for Concerted Research Actions. O.A. is a Research Associate of the F.R.S.-FNRS (Belgium). The early successes of this project wouldn't have been possible without the contributions of many collaborators outside the co-authors of this paper (aka the ``VORTEX team''). We would like to thank particularly P.~Riaud and J.~Baudrand for their decisive contributions to the genesis of this project, A.~Boccaletti and P.~Baudoz for maintaining and sharing with us their coronagraphic test bench at Observatoire de Paris-Meudon, E.~Pantin and the CEA team for inviting and helping us to piggyback on the VISIR upgrade, J.~Girard and the excellent ESO staff for making possible the installation of the AGPM on NACO, P.~Hinz and A.~Skemer for installing one of our L-band AGPMs on LMIRCam, B.~Mennesson and K.~Wallace for manufacturing optimized Lyot stops for the AGPM on LMIRCam, R.~Olivier for designing and manufacturing the opto-mechanical interfaces for the AGPM, D.~Vandormael for printing centering cross hairs on some of our AGPMs, J.-F.~Fagnard for cryo-testing the AGPMs, and finally C.~Hanot for his many contributions to this endeavor.

%%%%%%%%%%%%%%%%%%%%%%%%%%%%%%%%%%%%%%%%%%%%%%%%%%%%%%%%%%%%%
%%%%% References %%%%%

\bibliography{vortex_project}   %>>>> bibliography data in report.bib
\bibliographystyle{spiebib}   %>>>> makes bibtex use spiebib.bst

\end{document}